# A New Class of Electrically Tunable Metamaterial Terahertz Modulators

Rusen Yan, Berardi Sensale-Rodriguez, Lei Liu, Debdeep Jena and Huili Grace Xing[+]

Electrical Engineering Department, University of Notre Dame, Notre Dame, IN 46556, USA

[+] Email: hxing@nd.edu

**Switchable metamaterials offer unique solutions for efficiently manipulating electromagnetic waves, particularly for terahertz waves, which has been difficult since naturally occurring materials rarely respond to terahertz frequencies controllably. However, few terahertz modulators demonstrated to date exhibit simultaneously low attenuation and high modulation depth. Here we propose a new class of electrically-tunable terahertz metamaterial modulators employing metallic frequency-selective-surfaces (FSS) in conjunction with capacitively-tunable layers of electrons, promising near 100% modulation depth and < 15% attenuation. The fundamental departure in our design from the prior art is tuning enabled by self-gated electron layers, which is independent from the metallic FSS. Our proposal is applicable to all possible electrically tunable elements including graphene, Si, $MoS_2$ etc, thus opening up myriad opportunities for realizing high performance or flexible switchable metamaterials over an ultra-wide terahertz frequency range.**

Over the past decades the terahertz frequency regime (0.1 – 30 THz) has become the subject of much attention due to its important applications in astronomy, imaging, spectroscopy, etc.[1,2,3] At present, the developments of elements to efficiently control and manipulate THz waves are still lagging behind. In 2004 Kleine-Ostmann *et al.* proposed a semiconductor structure with metal-gated two-dimensional-electron-gas (2DEG) to modulate the intensity of terahertz waves with high insertion loss (~ 90% intensity loss) and poor modulation depths (~ 6%) for broadband operation at room temperature.[4,5] Alternatively, THz modulators based on metamatreials have gained a lot of attentions due to their moderate performance in modulation depth and attenuation loss. Recently our group have proposed and experimentally demonstrated a graphene-based THz amplitude modulators with dramatically minimized insertion loss of 5%.[6,7] However, all these attempts reported to data exhibit unavoidable trade-offs in high-quality performances including high modulation depth, low signal attenuation loss, polarization independence, facile fabrications and design flexibility. Here, we propose a new class of electrically tunable metamaterials consisting of capacitively-coupled layers of electrons that are structurally complementary to, but, only electromagnetically connected with the metallic FSS structures. Our proposed devices allow one to construct tunable metamaterials by placing the self-gated electron layers at optimal locations relative to the metallic FSS. This simple but elegant solution enables remarkably low insertion loss and high modulation depth, simultaneously. We also emphasize that the design methodology described in this letter can be readily extended to other switchable metamaterials based on either operation mode as well as all other electrically tunable materials including conventional semiconductors, transition metal oxides and so on.

Below, to illustrate this key design strategy, we use a pair of graphene as example self-gated electron layers and a square lattice of gold cross-slots as an example metallic FSS to construct a amplitude switchable metamaterial. An example of the proposed THz modulator is shown in Fig.1 (a). It is a tri-layer stack, consisting of a square lattice of gold cross-slot FSS and a pair of capacitively-coupled graphene layers situated at a distance $d$. The dielectric separating the FSS and graphene pair and between graphene layers is assumed to be $SiO_2$. In this structure, the gold-mesh FSS behaves as a band-pass filter with its center frequency ($f_0$) and bandwidth ($\Delta f$) determined by the mesh grid dimensions;[8] Graphene pairs plays the role of modulating the amplitude of the transmitted terahertz waves.[6,7] The energy band diagrams of this graphene pair are shown in Fig. 1 (b) where it can be seen that the transmission of THz waves reaches maximum when Fermi levels in the graphene pair are at Dirac point and decreases as Fermi levels are tuned further from Dirac pint. Considering the symmetric band structure of graphene around its Dirac point,[9] we can assume the electrical thus optical conductivity of the two graphene layers to be the same. Therefore, the conductivity value assumed in all the figures in this letter is that of a single graphene layer or half of the total conductivity of the 2DEG-pair.

To obtain terahertz transmission properties, the two port S parameters of the modulators in Fig. 1 (a) were simulated using a finite element based 3D electromagnetic solver: high frequency structure simulator (HFSS) by ANSYS, Inc.. In Fig. 2 (a) and (b) the amplitude of both $S_{21}$ (transmittance) and $S_{11}$ (reflectance) are plotted as a function of half of the 2DEG-pair conductivity, for $d = 10$ µm. The absorption of the terahertz wave intensity by the metamaterial modulator can be extracted accordingly and is shown in Fig. 2 (c). With the increase of graphene conductivity, both reflection and absorption in the 2DEG-pair increase, leading to the decrease of transmission. At moderate conductivities ( 2x(0.001-2) mS ), as is desirable, the transmittance are tuned in a large range while the resonant frequency does not shift. At very high conductivities, e.g. 2x5 mS, the absorption tends to decrease because of the increasing reflection. The maximum absorption by the 2DEG-pair is about 50%, consistent with what we discovered in broadband graphene terahertz modulators[7]. In addition, the phase of transmittance is plotted in Fig. 2 (d), where one can see that there is negligible phase change at the resonant frequency except at very high conductivities. The finite phase change away from the resonant frequency contributes minimally to the amplitude modulation.[10]

In Fig. 3, the power transmittances of the stack structure at resonance frequency as a function of half of the 2DEG conductivities are plotted with varied separating distance between FSS and 2DEG-pair. The red, pink and green curves correspond to different placements of the 2DEG-pair, $d = 1, 5, 10$ μm. For a given 2DEG-pair conductivity, the closer it is placed to the metallic FSS, the lower the intensity transmittance due to stronger near-field effects. For a 2DEG-pair conductivity of 2x0.001 mS, the amplitude transmittance reaches near 91%, i.e. an intensity insertion loss of 1 - $0.91^2$ ~ 17%. For a 2DEG-pair conductivity of 2x5 mS, the amplitude transmittance is < 20%; this corresponds to a modulation depth of the amplitude transmittance: (91-20)/91 ~ 78%, or of the intensity transmittance: ($91^2$-$20^2$)/$91^2$ ~ 95%. More careful examination revealed that the finite conductivity of gold[11] and the 2DEG-pair results in absorption of 12% and 2%, respectively; and the rest of the intensity insertion loss (3%) is due to reflection of the metamaterial stemming from the substrate refractive index and thickness.[7]

Particularly, the pink shaded region marks the typical tunable conductivity range of a single-layer graphene. For $d = 10$ μm and the typical conductivity tunable range of a single-layer graphene, the resultant intensity modulation depth is about (56-16)/56 ~ 72% with an insertion loss of ~ 44%. This is remarkable since the modulation depth by a suspended graphene pair is only about 20% although with a much lower insertion loss of < 5%.[7] When a graphene pair is placed at $d = 1$ μm, the resultant intensity modulation depth is about (14-1)/14 ~ 93% with an insertion loss of 86%.

A high modulation depth is attractive; however, too close a placement is not desirable since it introduces an extremely high insertion loss.[12] It is also worth noting that, the black curve (analytical results [6,7]) and blue circle points (from HFSS simulation) in Fig. 3 correspond to the intensity transmittance through a single layer graphene suspended in air and they match each other perfectly. This excellent agreement of analytical and HFSS solutions demonstrate the validity of our modeling.

Indeed, by placing the electron layers in a cavity, their absorbance/reflectance at a given conductivity can be augmented; the closer the electron layers are to the metallic FSS, the stronger the enhancement (neglecting the substrate effects[7]). Although graphene is promising in constructing such an electrically-tunable terahertz modulator due to its facile integration with other materials and low cost in large-scale production, its minimum conductivity[13] due to the zero bandgap can introduce an appreciable insertion loss (see Fig. 3). To further improve the modulator performance, it is necessary to adopt other tunable elements with lower optical conductivities for the terahertz frequency range of interest. In this regard, Si becomes a strong contender among all possible candidates due to its maturity. One can potentially use a thin membrane of $Si/SiO_2/Si$ as the self-gated electron layers. The major advantage of the Si-Si pair is that its conductivity can be tuned to be below $10^{-5}$ mS, where absorption by mobile electrons is practically zero. Considering a moderate conductivity range of $10^{-5}$ - 0.5 mS, one can see from Fig. 3 that a near unity modulation depth with an insertion loss of < 15% can be obtained. One can also stack multiple tunable elements to enhance modulation depth with a limited range of tunable conductivity as long as they do not introduce excessive insertion loss. It is also conceived that those structures can be readily fabricated employing 2D semiconductor crystals with non-zero bandgap, such as $MoS_2$ and $WS_2$, [14,15,16] for their easy integration and compatibility with flexible substrates.

Finally we discuss the operation frequency range of the proposed switchable terahertz metamaterials enabled by the self-gated electron layers.   The optical conductivity of mobile electrons in the terahertz to infrared region (below photon energies that trigger interband transitions) is determined by their Drude and plasmonic behavior.   For simplicity, let us consider the Drude conductivity only: it decreases with increasing frequency.   For instance, the Drude dominated terahertz bandwidth in graphene has been found to be near 4 THz[7,17].   As a result, we can assume that the graphene optical and DC electrical conductivities are the same in the above example device designed near 1.3 THz.   However, another unique advantage of the proposed device in this letter is that it allows the location of the self-gated electron layers to be optimized depending on their tunable conductivity range at the designed terahertz frequency. For a graphene-based device for a frequency band < 4 THz, the optimal location is near $d$ =10  μm; while for a device for a band near 10 THz, the optimal location could be near $d$ = 1  μm. It is also straightforward to extend similar design considerations to devices employing tunable plasmonic effects, which play important roles above the Drude dominated frequency range.[17,18]  We also like to point out that our discussions focus on tunable elements, including graphene and other materials with tunable optical conductivity in the terahertz band, while the primary frequency selective features in the switchable metamaterials should be made of low-loss materials such as gold[11], silver[11] or dielectric cavities[19]  to minimize the device insertion loss.

In conclusion, we have proposed a new class of matermaterial-based tunable THz modulators with significant improvements in modulation depth, insertion loss, fabrication, integration and design flexibility. Our simple but unique idea ensures that the amplitude-modulating 2DEG-pair work independently with frequency-selective metamaterial, thus providing us much more freedom in the device design. Our proposal can be readily applied into a variety of promising materials, such as graphene, Si, $MoS_2$ etc, which holds promise for future experimental realization of high-performance tunable THz modulators.

ACKNOWLEDGMENTS Xing acknowledges the support from National Science Foundation (CAREER award). Jena and Xing acknowledges the support from National Science Foundation (ECCS-0802125) and from the Office of Naval Research (N00014-09-1-0639 and N00014-11-1-0721). Liu and Xing acknowledge the support from National Science Foundation (ECCS-1002088 and ECCS-1202452). Liu, Jena and Xing also acknowledge the support from the Center of Advanced Diagnostics & Therapeutics (AD&T), the Midwest Institute of Nanoelectronics Discovery (MIND) and the Center for Nanoscience and Technology (ND nano) at the University of Notre Dame.

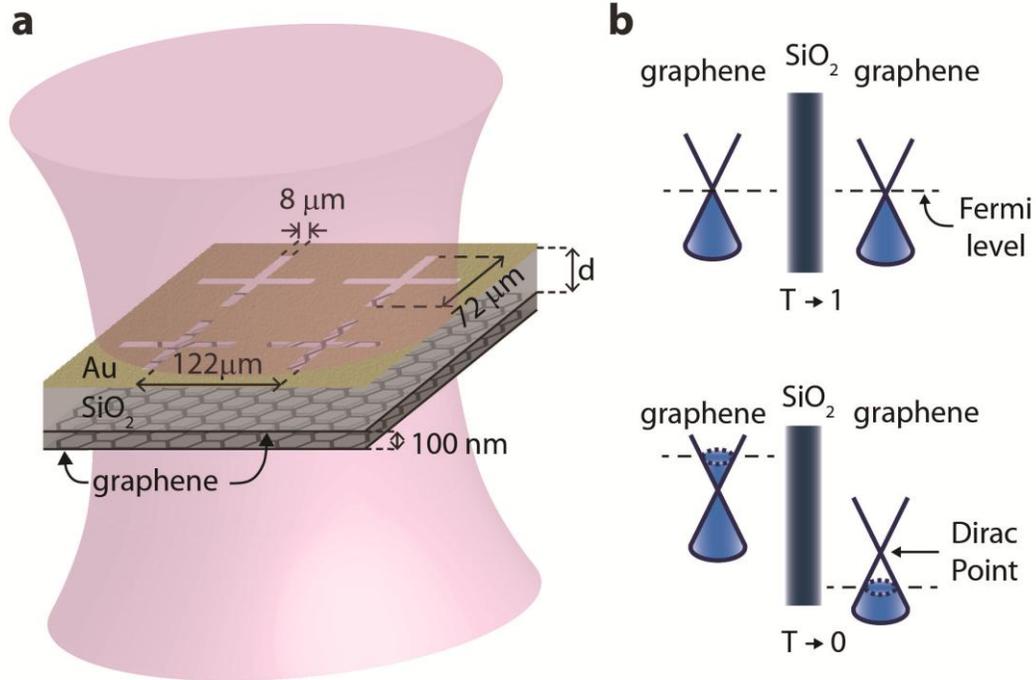

**Figure 1** a) The proposed THz modulator structure, consists of a square lattice of gold cross-slot FSS and a pair of capacitively-coupled graphene layers situated at a distance *d*. The dielectric separating the FSS and graphene pair and between graphene layers is assumed to be $SiO_2$. b) Energy band diagrams of the graphene pair. At zero bias, the Fermi level is at the Dirac point of both graphene layers, leading to minimum conductivity and maximum terahertz transmission at the resonant frequency. When biased, the Fermi level moves into the conduction and valence band of the two graphene layers, respectively, resulting in an enhanced conductivity and minimized terahertz transmission.

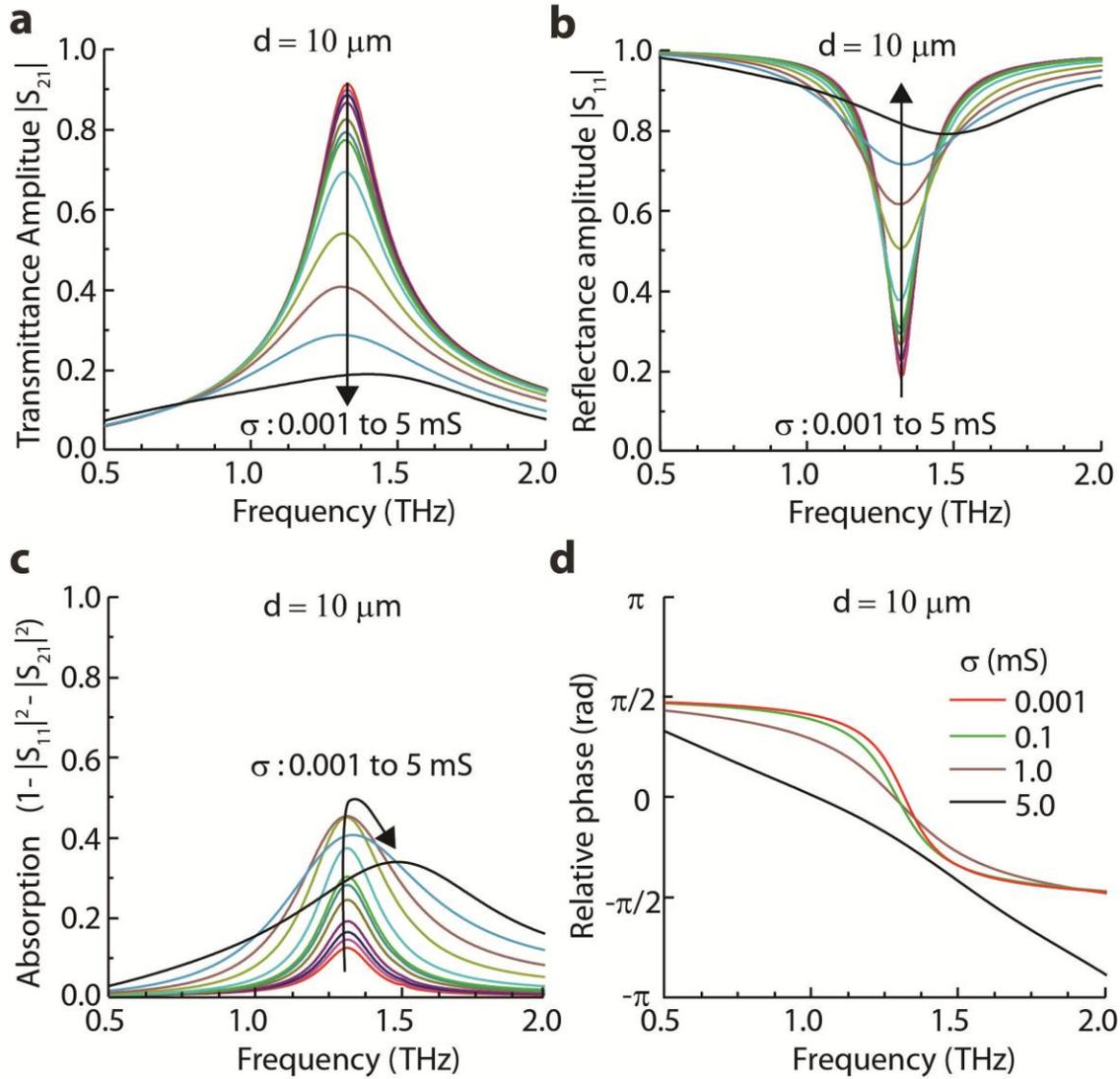

Fig. 2 (a) and (b) Amplitude of simulated transmittance ($S_{21}$) and reflectance ($S_{11}$) as a function of half of the 2DEG-pair conductivity.  c) Associated absorption of terahertz wave intensity.  d) Phase of transmittance at varied conductivities.

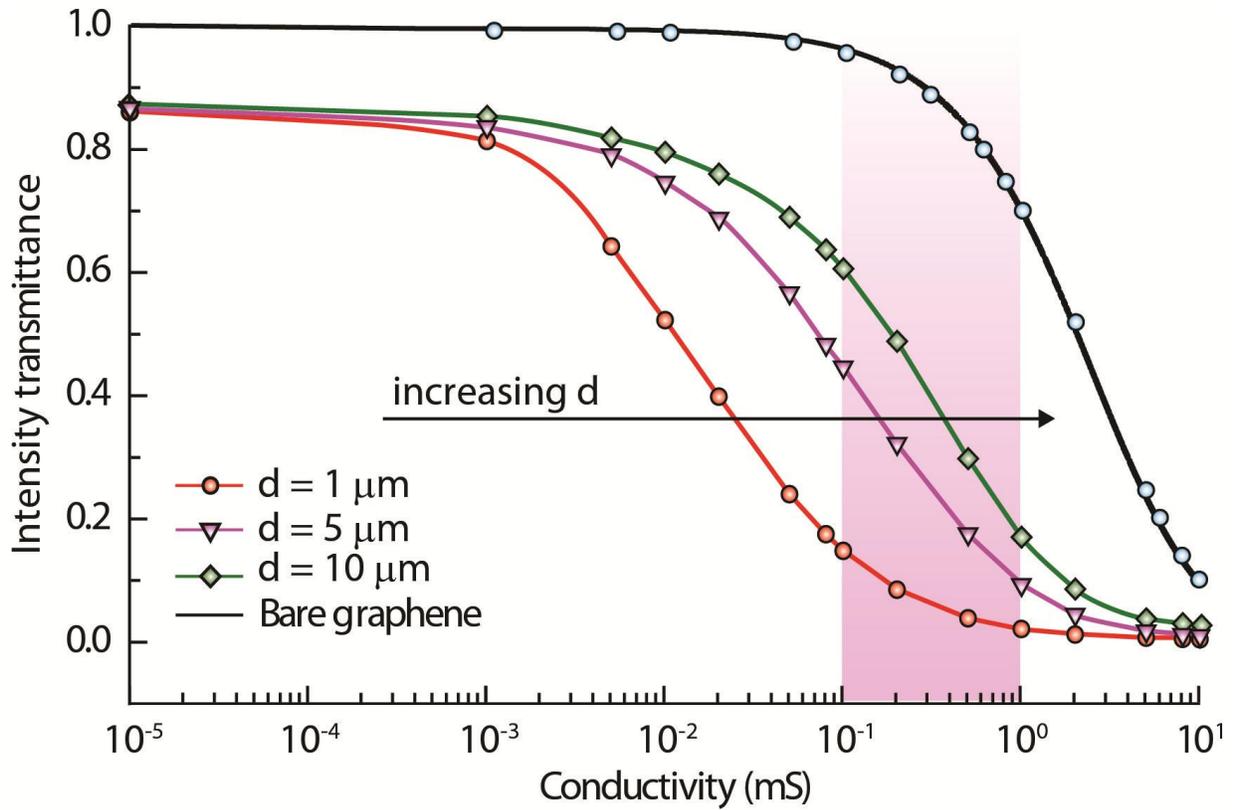

**Fig 3** The red, pink and green curves correspond to the intensity transmittance versus the half conductivity of the 2DEG-pair at different separating distance between the pair and metamaterial: $d = 1, 5, 10$ μm. The black curve and blue circles points are respectively analytical solutions[7] and HFSS simulation results corresponding to the intensity transmittance through a single layer graphene suspended in air. The pink shaded region shows the typical range of single-layer graphene conductivity.

**References:**

[1] A.D. Boardman, V.V. Grimalsky, Y.S. Kivshar, S.V. Koshevaya, M. Lapine, N.M. Litchinitser, V.N. Malnev, M. Noginov, Y.G. Rapoport, V.M. Shalaev, *Laser Photon. Rev.* **5** (2), 287-307 (2010).

[2] M. Tonouchi, *Nat. Photon.* **1**, 97 - 105 (2007).

[3] H. T. Chen, W. J. Padilla, J. M. O. Zide, A. C. Gossard, A. J. Taylor and R. D. Averitt, *Nature* **444**, 597-600 (2006).

[4] T. Kleine-Ostmann, T, P. Dawson, K. Pierz, G. Hein and M. Koch, *App. Phys. Lett.* **84** (18), 3555-3557 (2004).

[5] T. Kleine-Ostmann, K. Pierz, G. Hein, P. Dawson, M. Marso and M. Koch, *J. Appl. Phys.* **105**, 093707 (2009).

[6] Berardi Sensale-Rodriguez, Tian Fang, Rusen Yan, Michelle M. Kelly, Debdeep Jena, Lei Liu, and Huili (Grace) Xing, Appl. Phys. Lett. 99, 113104 (2011).

[7] B. Sensale-Rodriguez, R. Yan, M. M. Kelly, T. Fang, K. Tahy, W. S. Hwang, D. Jena, L. Liu and H. G. Xing , Nature Communications **3** 780 (2012).

[8] D. W. Porterfield, J. L. Hesler, R. Densing, E. R. Mueller, T. W. Crowe and R. M. Weikle, *Appl. Opt.* **33**(25), 6042-6052 (1994).

[9] K. S. Novoselov, A. K. Geim, S. V. Morozov, D. Jiang, Y. Zhang, S. V. Dubonos, I. V. Grigorieva and A. A. Firsov, *Science,* **306**, 666 (2004).

[10] H.T. Chen, W. J. Padilla, M. J. Cich, A. K. Azad, R. D. Averitt and A. J. Taylor. *Nat. Photon.* **3**, 148-151 (2009).

[11] P. Tassin, T. Koschny, M. Kafesaki and C. M. Soukoulis, *Nat. Photon.* **6**, 259-264 (2012).

[12] S. H. Lee, M. Choi, T. T. Kim, S. Lee, M. Liu, X. Yin, H. K. Choi, S. S. Lee, C.G. Choi, S.Y. Choi, X. Zhang, B. Min, *Switching terahertz waves with gate-controlled active graphene metamaterials.* Preprint at http://arxiv.org/abs/1203.0743 (2012).

[13] Y. W. Tan, Y. Zhang, K. Bolotin1, Y. Zhao, S. Adam, E. H. Hwang, S. Das Sarma, H. L. Stormer, and P. Kim, *Phys. Rev. Lett.* **99**, 246803 (2007).

[14] R. Radisavljevic, A. Radenovic, A. Giacometti and A. Kis, *Nat. Nanotech.,* **6**, 147-150 (2010).

[15] M. F. Mak, C. Le, J. Hone, J. Shan and T. F. Heinz, T. F, *Phys. Rev. Lett.* **105**, 136805 (2010).

[16] W. S. Hwang, M. Remskar, R. Yan, V. Protasenko, K. Tahy, S. D. Chae, P. Zhao, A. Konar, H. G. Xing, A. Seabaugh and D. Jena, *Appl. Phys. Lett.* **101**, 013107 (2012)

[17] H. T. Chen, J. F. O'Hara, A. K. Azad, A. J. Taylor, R. D. Averitt, D. B. Shrekenhamer and W. J. Padilla, *Nat. Photon.* **2**, 295-298 (2008).

[18] S. J. Allen, D. C. Tsui and R. A. Logan, *Phys. Rev. Lett.* **38**, 980-983 (1977).

[19] M. Liu, X. Yin, E. Ulin-Avila, B. Geng, T. Zentgraf, L. Ju, F. Wang and X. Zhang, *Nature* **474**, 64-67 (2011).